\documentclass[
 aps, pra,
 amsmath,amssymb,
 10pt,
 final,
tightenlines,
 twoside,
 twocolumn,
 nobibnotes,
nofootinbib,
 superscriptaddress,
showkeys,
 ]
{revtex4-2}

\usepackage[utf8x]{inputenc}
\usepackage[english]{babel}
\usepackage{graphicx}% Include figure files
\usepackage{dcolumn}% Align table columns on decimal point
\usepackage{bm}% bold math
\usepackage{natbib}
\usepackage{float}
\usepackage{hyperref}
\usepackage{slashed}
\usepackage{multirow}
\usepackage{lineno}
\usepackage{booktabs}
\usepackage{silence}
\usepackage{xcolor}
\usepackage{mathtools}

\begin{document}

\selectlanguage{english}

%\keywords{stars: double or binary---stars: individual: ADS\,48}

%\ydk{}
%\titlerunning{}
%\authorrunning{}
%\toctitle{}
%\tocauthor{}

\title{Top anomalous chromomagnetic dipole moment in a two Higgs doublet model}

\author{\firstname{T.}~\surname{Cisneros-Pérez} }
 \email{tzihue@gmail.com}
  \affiliation{Unidad Académica de Ciencias Químicas, Universidad Autónoma de  Zacatecas,Apartado Postal C-585, 98060 Zacatecas, M\'exico. }

\author{\firstname{A.}~\surname{Ramirez-Morales}}
 \email{andres.ramirez@fisica.uaz.edu.mx}
\affiliation{Facultad de Física, Universidad Autónoma de Zacatecas, Apartado Postal C-580, 98060 Zacatecas, M\'exico}

\author{\firstname{J.}~\surname{Montaño-Domínguez}}
 \email{jmontano@conahcyt.mx}
\affiliation{Facultad de Ciencias Físico Matemáticas, Universidad Michoacana de San Nicolás de Hidalgo,
Av. Francisco J. Múgica s/n, C. P. 58060, Morelia, Michoacán, México.}

\author{\firstname{M. A.}~\surname{Hernández-Ruíz}}

\author{\firstname{A.}~\surname{Gutiérrez-Rodríguez}}

\begin{abstract}
We review the anomalous Chromomagnetic Dipole Moment (CMDM) of the top quark in a Two Higgs Doublet Model (2HDM). We include interactions with the involvement of the extended CKM matrix in this model, we obtain new bonds for the future experimental measurements and theoretical considerations.

\end{abstract}

\maketitle

\section{INTRODUCTION}

 We know of the exceptional properties of the top quark to explore its CMDM in the 2HDM developed by \cite{schmaltz2010bestest} and update its magnitude by including contributions arising from the new interactions~\cite{Cisneros-Perez2023}  with flavor-changing mediated by two matrices $V_{Hu}$ and $V_{Hd}$ related with the $V_{CKM}$ matrix (Cabibbo-Kobayashi-Maskawa) \cite{kobayashi1973cp}.
In this context, the theoretical framework 2HDM offers an enriched perspective within the broader landscape of Little Higgs Models (LHM). It concisely and naturally addresses issues such as custodial symmetry violation, divergent singlets and others \cite{schmaltz2010bestest}. The principal aspect of this model is its modular structure, which requires two distinct breaking scales, $f$ and $F$, where $F>f$.  Despite the advantages and solutions offered by the model developed in \cite{schmaltz2010bestest}, it has not been explored as extensively as other models within the LHM family.

Currently, the measurement of the CMDM given by the CMS Collaboration at the LHC \cite{Sirunyan2020} using $pp$ collisions at $13$ TeV and  with an integrated luminosity of $35.9$ fb$^{-1}$ is

\begin{equation}\label{cmdmexp}
    \hat{\mu}^{Exp}_t=-0.024^{+0.013}_{-0.009}(\mathrm{stat})^{+0.016}_{-0.011}(\mathrm{syst}).
\end{equation}

 In the case of 2HDM models, the calculation of the CMDM has been performed in the Littlest Higgs with T symmetry (LHT) \cite{Ding:2008nh} with an estimate of $10^{-3}-10^{-4}$ and in four generations of fermions \cite{Hernandez-Juarez:2018uow}  with the result $10^{-1}-10^{-2}$.
 In our case, we have calculated the $\hat{\mu}_t^{BLHM}$ for each angle $\beta$ in the range $0.15\leq\beta\leq1.49$ with results $10^{-4}\leq\hat{\mu}_t^{BLHM}\leq10^{-2}$.

This paper is structured as follows: In Section~\ref{review}, we provide a brief review of the 2HDM in order to establish the physics frameworks. In Section~\ref{chromo}, we describe the sector of the BLHM that includes $gt\bar{t}$ vertex needed for the $\hat{\mu}$ calculation. Section~\ref{pspace}, discusses the allowed parameter space within the calculations of the CMDM. In Section~\ref{pheno}, we detail the  scenarios in which we calculate the CMDM. Section~\ref{results} contains  our results. We give our conclusions in Section~\ref{conclusions}.

\section{2HDM insights}
\label{review}

The model~\cite{schmaltz2010bestest} is constructed from the symmetry group $SO(6)_A \times SO(6)_B$, which breaks and adquires a vacuum expectation value (VEV). This process gives rise to 15 pseudo-Nambu-Goldstone bosons.

%
%%
%%%
%%%% SECTOR ESCALAR
%%%
%%
%

\subsection{Structure of the Scalar Sector}
In this 2HDM, we can write the quartic potential as~\cite{schmaltz2010bestest}

\begin{equation}\label{potVq}
 V_q=\frac{1}{4}\lambda_{65}f^4(\Sigma_{65})^2+\frac{1}{4}\lambda_{56}f^4(\Sigma_{56})^2,
\end{equation}

\noindent where $\lambda_{56}$ and $\lambda_{65}$ are coefficients that must be nonzero to achieve collective symmetry breaking. After an expansion and using Eq. (\ref{potVq}), we obtain

\begin{eqnarray}\label{potVqSerie}
 V_q=\frac{\lambda_{65}}{2}\left(f\sigma-\frac{1}{\sqrt{2}}h_1^Th_2+\dots\right)^2\\\nonumber
 +\frac{\lambda_{56}}{2}\left(f\sigma+\frac{1}{\sqrt{2}}h_1^Th_2+\dots\right)^2.
\end{eqnarray}

Together, the two terms in Eq.(\ref{potVqSerie}) generate a tree-level quartic potential for the Higgs after integrating out $\sigma$\cite{schmaltz2010bestest,Schmaltz:2008vd,phdthesis}:

\begin{equation}
 V_q=\frac{1}{2}\lambda_{0}\left(h_1^Th_2 \right)^2.
\end{equation}

\noindent In this way, we derive a quartic collective potential that is proportional to two distinct couplings~\cite{schmaltz2010bestest}. It is evident that $\lambda_0$ vanishes when either $\lambda_{56}$, $\lambda_{65}$, or both are zero.

\noindent Without gauge interactions, not all scalar fields acquire mass; hence, it is necessary to introduce the potential,
\begin{eqnarray}\label{potVs}\nonumber
 V_s&=&-\frac{f^2}{4}m_4^2Tr\left(\Delta^{\dagger}M_{26}\Sigma M^{\dagger}_{26}+\Delta M_{26}\Sigma^{\dagger}M^{\dagger}_{26}\right)\\
 &&-\frac{f^2}{4}\left(m_5^2\Sigma_{55}+m_6^2\Sigma_{66}\right),
\end{eqnarray}

\noindent Here, $m_4$, $m_5$, and $m_6$ represent mass parameters, while $(\Sigma_{55}, \Sigma_{66})$ are matrix components.

\noindent The operator $\Delta$ originates from a global symmetry $SU(2)_C \times SU(2)_D$. It can be parameterized as
\begin{equation}
 \Delta=e^{2i\Pi_d/F},\hspace{0.3cm}\Pi_d=\chi_a\frac{\tau_a}{2}\hspace{0.3cm}(a=1,2,3),
\end{equation}
 The matrix $\Pi_d$ includes the scalars from the triplet $\chi_a$, which mix with the triplet $\phi_a$, while $\tau_a$ denotes the Pauli matrices.
If we expand the operator $\Delta$ in powers of $1/F$ and substitute it into Eq. (\ref{potVs}),
\begin{equation}
 V_s=\frac{1}{2}\left(m^2_{\phi}\phi^2_a+m^2_{\eta}\eta^2_a+m^2_1h^T_1h_1+m^2_2h^T_2h_2\right),
\end{equation}
where
\begin{eqnarray}
 m^2_{\phi}&=&m^2_{\eta}=m^2_4,\\\nonumber
 m^2_1&=&\frac{1}{2}(m^2_4+m^2_5),\\\nonumber
 m^2_2&=&\frac{1}{2}(m^2_4+m^2_6).
\end{eqnarray}
To have an EWSB, the next potential term is introduced\cite{schmaltz2010bestest}:
\begin{equation}
 V_{B_{\mu}}=m_{56}^2f^2\Sigma_{56}+m_{65}^2f^2\Sigma_{65},
\end{equation}

In this manner, we have the complete scalar potential,
\begin{equation}\label{pEscalar}
 V=V_q+V_s+V_{B_{\mu}}.
\end{equation}

\noindent We need a potential to the Higgs doublets thus, we minimize Eq. (\ref{pEscalar}) with respect to $\sigma$ and then substitute the resulting expression back into Eq. (\ref{pEscalar}),
\begin{eqnarray}\label{potVH}\nonumber
 V_H&=&\frac{1}{2}\Big[m_1^2h_1^Th_1+m_2^2h_2^Th_2-2B_{\mu}h_1^Th_2\\
 &+&\lambda_0(h_1^Th_2)^2\Big],
\end{eqnarray}
where

\begin{equation}\label{potB}
 B_{\mu}=2\frac{\lambda_{56}m_{65}^2+\lambda_{65}m_{56}^2}{\lambda_{56}+\lambda_{65}}.
\end{equation}
\noindent The potential (\ref{potVH}) reaches a minimum when $m_1m_2 > 0$. After EWSB, the Higgs doublets obtain VEVs,
\begin{equation}\label{aches}
 \langle h_1\rangle=v_1,\hspace{0.3cm}\langle h_2\rangle=v_2.
\end{equation}

\noindent The two terms in (\ref{aches}) must be minimized in Eq. (\ref{potVH}),

\begin{eqnarray}
 v_1^2=\frac{1}{\lambda_0}\frac{m_2}{m_1}(B_{\mu}-m_1m_2),\\
 v_2^2=\frac{1}{\lambda_0}\frac{m_1}{m_2}(B_{\mu}-m_1m_2).
\end{eqnarray}

The angle $\beta$ between $v_1$ and $v_2$ is defined, such that:

\begin{equation}
\tan\beta=\frac{\langle h_{11}\rangle}{\langle h_{21}\rangle}=\frac{v_1}{v_2}=\frac{m_2}{m_1},
\end{equation}

in this way, we have

\begin{eqnarray}
v^2&=&v_1^2+v_2^2\\\nonumber
&=&\frac{1}{\lambda_0}\left(\frac{m_1^2+m_2^2}{m_1m_2}\right)(B_{\mu}-m_1m_2)\\\nonumber
&\simeq&(246\;GeV)^2.
\end{eqnarray}

After the EWSB, the scalar sector gives rise to massive states such as $h^0$, $A^0$, $H^{\pm}$, and $H^0$, with their masses:
\begin{eqnarray}
\label{masa-esc1}
&&m^2_{G^0}=m^2_{G^{\pm}}=0,\\
\label{masa-esc2}
&&m^2_{A^0}=m^2_{H^{\pm}}=m^2_1+m^2_2,\label{masa-A0}\\
\label{masa-esc3}
&&m^2_{H^0}=\frac{B_{\mu}}{\sin2\beta}\label{masa-H0}\\\nonumber
+&&\sqrt{\frac{B^2_{\mu}}{\sin^22\beta}-2\lambda_0\beta_{\mu} v^2\sin2\beta+\lambda_0^2v^4\sin^22\beta},
\end{eqnarray}
where $G^0$ and $G^{\pm}$ are Goldstone bosons that are eaten to give masses to the $W^{\pm}, Z$ bosons of the SM.

%                     %
%%                   %%
%%% SECTOR DE NORMA %%%
%%                   %%
%                     %

\subsection{Gauge Boson Sector}

The following Lagrangian provides the gauge kinetic terms in this model,
\begin{equation}\label{lag-norma}
\mathcal{L}=\frac{f^2}{8}Tr\left(D_{\mu}\Sigma^{\dag}D^{\mu}\Sigma\right)+\frac{F^2}{4}Tr\left(D_{\mu}\Delta^{\dag}D^{\mu}\Delta\right),
\end{equation}
 \( D_{\mu} \Sigma \) and \( D_{\mu} \Delta \) are covariant derivatives.

These are connected to the couplings \( g_A \) and \( g_B \) of \( SU(2)_A \times SU(2)_B \),

\begin{eqnarray}\label{acoplesSU}
g&&=\frac{g_Ag_B}{\sqrt{g_A^2+g_B^2}},\\
s_g&&=\sin\theta_g=\frac{g_A}{\sqrt{g_A^2+g_B^2}},\\
c_g&&=\cos\theta_g=\frac{g_B}{\sqrt{g_A^2+g_B^2}},
\end{eqnarray}
here, $\theta_g$ is the mixing angle.

The masses of the heavy gauge bosons \( W^{\prime \pm} \), \( Z' \), as well as the masses of the SM bosons, are also generated~\citep{schmaltz2010bestest, phdthesis},
\begin{eqnarray}
\label{masa-zp}
m_{Z'}^2&=&\frac{1}{4}(g_A^2+g_B^2)(f^2+F^2)-\frac{1}{4}g^2v^2,\\\nonumber
\label{masa-wp}
m_{W'}^2&=&\frac{1}{4}(g_A^2+g_B^2)(f^2+F^2)-m_W^2.
\end{eqnarray}

%                    %
%%                  %%
%%% SECTOR FERMION %%%
%%                  %%
%                    %

\subsection{Fermion sector}

The structure of the fermion sector is given by the Lagrangian~\cite{schmaltz2010bestest}
\begin{eqnarray}
\label{lag-yuk}
\mathcal{L}_t&=&y_1fQ^TS\,\Sigma\, SU^c+y_2fQ_a^{\prime T}\Sigma\,U^c\\\nonumber
&+&y_3fQ^T\Sigma\, U_5^{\prime c}+y_bfq_3^T(-2iT_ R^3\Sigma)U_b^c+\textrm{h.c.},
\end{eqnarray}

\noindent where $(Q,Q')$ and $(U,U')$ are multiplets of $SO(6)_A$ and $SO(6)_B$, respectively,~\cite{phdthesis}.
$S=diag(1,1,1,1,-1,-1)$ is a symmetry operator, $(y_1,y_2,y_3)$ represent Yukawa couplings, and the term $(q_3,U_b^c)$ in Eq. (\ref{lag-yuk}) contains information about the bottom quark.
The  new heavy quarks in the 2HDM framework play a key role in addressing the hierarchy problem. Those heavy quarks are: $T$, $T^5$, $T^6$, $T^{2/3}$, $T^{5/3}$, and $B$,~\cite{schmaltz2010bestest}.
In the quark sector Lagrangian, the Yukawa couplings must satisfy $0<y_i<1$. The expresion for the quark top mass contains its Yukawa coupling $y_t$~\cite{phdthesis}, such that:
\begin{equation}
m^2_t=y_t^2v_1^2\label{masa-top}.
\end{equation}

\noindent The coupling $y_t$ have the form:
\begin{equation}\label{acople-yt}
 y_t^2=\frac{9y_1^2y_2^2y_3^2}{(y_1^2+y_2^2)(y_1^2+y_3^2)},
\end{equation}
that is part of the measure of fine-tuning in the BLHM \cite{phdthesis},
\begin{equation}
\label{ajuste-fino}
\Psi=\frac{27f^2}{8\pi^2v^2\lambda_0\cos^2\beta}\frac{|y_1|^2|y_2|^2|y_3|^2}{|y_2|^2-|y_3|^2}\log\frac{|y_1|^2+|y_2|^2}{|y_1|^2+|y_3|^2}.
\end{equation}

\subsection{Flavor Mixing in the 2HDM}

We take the flavor structure proposed in \cite{Cisneros-Perez2023}, as it enhances the number of interactions. The modification add terms to the Lagrangians that describe interactions between the fields $(W^{\prime\pm},H^{\pm},\phi^{\pm},\eta^{\pm})$, the heavy quark $B$, and the light SM quarks $(u, c, d, s)$.

These made contributions to the 2HDM in \cite{Cisneros-Perez2023} to the phenomenology provided by the model in such a way that two CKM unitary matrices, $V_{Hu}$ and $V_{Hd}$, can now be associated, satisfying the relation $V_{CKM}=V_{Hu}^{\dagger}V_{Hd}$, where $V_{CKM}$ is the Cabibbo-Kobayashi-Maskawa matrix \cite{kobayashi1973cp}.

\section{The chromomagnetic dipole moment}
\label{chromo}

The description of the contributions of the vertex $g\bar{t}t$ is given by the effective Lagrangian:

\begin{equation}
    \mathcal{L}_{eff}=-\frac{1}{2}\bar{t}\sigma^{\mu\nu}\left(\hat{\mu}_t+i\hat{d}_t\gamma^5\right)tG^a_{\mu\nu}T^a,
\end{equation}

\noindent where $G^a_{\mu\nu}$ is the gluon strength tensor, $T^a$ are the $SU(3)$ generators, $\hat{\mu}_t$ is the CMDM and $\hat{d}_t$ is the CEDM such that

\begin{equation}\label{dip-usual}
    \hat{\mu}_t=\frac{m_t}{g_s}\mu_t,\hspace{0.5cm}\hat{d}_t=\frac{m_t}{g_s}d_t.
\end{equation}
The definitions given for Eq.~(\ref{dip-usual}) are the standard relations for the CMDM and the CEDM in the literature on the subject because $\mathcal{L}_{eff}$ has dimension 5, where $m_t$ is the mass of the top quark and $g_s=\sqrt{4\pi \alpha_s}$ is the coupling constant of the group.

\section{Parameter space of the BLHM}
\label{pspace}

We take the full parameter space from \cite{Cisneros-Perez2023} and we show it in the parameter Tables \ref{masas-esc}, \ref{masas-esc1}, \ref{masas-boson} and \ref{masas-quarks}.

\begin{table}
\caption {Parameters and scalar masses of the BLHM.} \label{masas-esc}
\medskip
\begin{tabular}{c|c|c|c}
\hline
\multirow{2}{*}{Parameter } & \multicolumn{3}{c}{Values}  \\
\cline{2-4}
  & Min & Max & Unit \\
\hline
$\beta$                          & 0.15    & 1.49       & rad \\
$\alpha$                         & -0.25   & 1.08       & rad \\
$y_3$                            & 0.32    & 0.96       & --  \\
$m_{A^0}$                        & 125.00  & 1693.04    & GeV \\
$m_{H^0}$                        & 872.04  & 1900.3     & GeV \\
$m_{H^{\pm}}$                    & 125.00  & 1693.04    & GeV \\
\hline
\end{tabular}
\end{table}

\begin{table*}
\caption {Parameters and scalar masses  constrained in the BLHM.} \label{masas-esc1}
\medskip
\begin{tabular}{c|c|c|c|c|c|c|c}
\hline
\multirow{2}{*}{Parameter } & \multicolumn{2}{c}{$f(1\,{\scriptstyle\text{TeV}})$}  & \multicolumn{2}{c}{$f(2\,{\scriptstyle\text{TeV}})$} & \multicolumn{2}{c}{$f(3\,{\scriptstyle\text{TeV}})$}&\\
\cline{2-8}
  & \hspace{3mm}Min\hspace{3mm} & \hspace{3mm}Max\hspace{3mm} & \hspace{3mm}Min\hspace{3mm} & \hspace{3mm}Max\hspace{3mm} & \hspace{3mm}Min\hspace{3mm}  & \hspace{3mm}Max\hspace{3mm} & \hspace{3mm}Unit\hspace{3mm}  \\
\hline
$\beta$                          & 0.79    & 1.47       & 0.79   & 1.36  &  0.79 & 1.24 & rad\\
$\alpha$                         & 0.38    & 1.06       & 0.38   & 0.95  &  0.38 & 0.83 & rad\\
$\Psi$                           & 0.096   & 2.11       & 0.38   & 2.03  & 0.87  & 2.04 & -- \\
$m_{A^0}$                        & 125.0  & 884.86     & 125.0   & 322.75   & 125.0  & 207.07  & GeV\\
$m_{H^0}$                        & 872.04  & 1236.06     & 872.04   & 921.42  & 872.04  & 887.53  & GeV\\
$m_{H^{\pm}}$                    & 125.0  & 884.86     & 125.0   & 322.75   & 125.0  & 207.07  & GeV\\
\hline
\end{tabular}
\end{table*}

\begin{table}[H]
\begin{center}
\caption {Scalar masses of the BLHM. } \label{masas-boson}
\medskip
\begin{tabular}{c|c|c|c}
\hline
\multirow{2}{*}{Mass\hspace{3mm}} & \multicolumn{2}{c}{Values} & \\
\cline{2-4}
  & $f(1\,{\scriptstyle\text{TeV}})$ & $f(3\,{\scriptstyle\text{TeV}})$  & Unit\\
\hline
$m_{\sigma}$                          & 1414.2    & 4242.6    & GeV    \\
$m_{\phi^0}$                          & 836.1    & 999.3    & GeV    \\
$m_{\phi^{\pm}}$                         & 841.9   & 1031.9      & GeV  \\
$m_{\eta^{\pm}}$                            & 580.0    & 1013.9      & GeV   \\
\hline
\end{tabular}
\end{center}
\end{table}

\begin{table}[H]
\begin{center}
\caption {Quarks masses of the BLHM. } \label{masas-quarks}
\medskip
\begin{tabular}{c|c|c|c}
\hline
\multirow{2}{*}{Mass\hspace{3mm}} & \multicolumn{2}{c}{\hspace{3mm}$1\leq f\leq 3$ {\scriptsize TeV}} & \\
\cline{2-4}
  & Min & Max  & Unit\\
\hline
$m_{T}$                          & 1140.18    & 3420.53    & GeV    \\
$m_{T^5}$                          & 773.88    & 2321.66    & GeV    \\
$m_{T^6}$                         & 780.0   & 2100.0      & GeV  \\
$m_{T^{2/3}}$                            & 780.0    & 2100.0      & GeV   \\
$m_{T^{5/3}}$                            & 780.0    & 2100.0      & GeV   \\
$m_{B}$                            & 1140.18    & 3420.53      & GeV   \\
\hline
\end{tabular}
\end{center}
\end{table}

\section{Phenomenology of the CMDM}
\label{pheno}

In the context of the CMDM, we calculate one loop diagram with the new heavy scalar and gauge bosons and a second one loop diagram with the new heavy fermions. The amplitudes are given by:
\begin{eqnarray}\label{ampE}
  &&\mathcal{M}^{\mu}_t(S_i)=\sum_{j}\int\frac{d^4k}{(2\pi)^4}\bar{u}(p^{\prime})(S^{\ast}_i+P^{\ast}_i\gamma^5)\delta_{A\alpha_1}\\\nonumber
  &&\times\left[i\frac{\slashed{k}+\slashed{p}^{\prime}+m_{Q_j}}{(k+p^{\prime})^2-m_{Q_j}^2}\delta_{\alpha_1\alpha_3}\right]\left(-ig_s\gamma^{\mu}T^a_{\alpha_2\alpha_3}\right)\\\nonumber
  &&\times\left[i\frac{\slashed{k}+\slashed{p}+m_{Q_j}}{(k+p)^2-m^2_{Q_j}}\right](S_i+P_i\gamma^5)\delta_{B\alpha_4}V_{Ht}^{\ast}V_{Ht}u(p)\\\nonumber
  &&\times\left(\frac{i}{k^2-m^2_{S_i}}\right),
\end{eqnarray}
\noindent and
\begin{eqnarray}\label{ampV}
  &&\mathcal{M}^{\mu}_t(V_i)=\sum_{j}\int\frac{d^4k}{(2\pi)^4}\bar{u}(p^{\prime})\gamma^{a_1}(V^{\ast}_i+A^{\ast}_i\gamma^5)\delta_{A\alpha_1}\\\nonumber
  &&\times\left[i\frac{\slashed{k}+\slashed{p}^{\prime}+m_{Q_j}}{(k+p^{\prime})^2-m_{Q_j}^2}\delta_{\alpha_1\alpha_3}\right]\left(-ig_s\gamma^{\mu}T^a_{\alpha_2\alpha_3}\right)\\\nonumber
  &&\times\left[i\frac{\slashed{k}+\slashed{p}+m_{Q_j}}{(k+p)^2-m^2_{Q_j}}\right]\gamma^{a_2}(V_i+A_i\gamma^5)\delta_{B\alpha_4}V^{\ast}_{Ht}V_{Ht}u(p)\\\nonumber
  &&\times\left[\frac{i}{k^2-m^2_{V_i}}\left(-g_{\alpha_1\alpha_2}+\frac{k_{\alpha_1}k_{\alpha_2}}{m^2_{V_i}}\right)\right],
\end{eqnarray}

\noindent where $T^a_{\alpha_n\alpha_m}$ are the $SU(3)$ generators.  The coefficients $(S_j,P_j,V_j,A_j)$ carry all contributions from the 2HDM quantified by the vertices $\bar{Q}_jS_it$, $\bar{t}S_i^{\dagger}Q_j$ for scalar and pseudoscalar interactions, and $\bar{Q}_jV_it$, $\bar{t}V_i^{\dagger}Q_j$ for vector and axial interactions, respectively. We used the \texttt{FeynCalc} package \cite{shtabovenko2020feyncalc} and the \texttt{Package X} \cite{patel2015package} for \texttt{Mathematica} in this calculations.

In interactions with charged bosons it is necessary to consider the extended CKM matrix for the 2HDM, $V_{CKM}=V_{Hu}^{\dagger}V_{Hd}$, introduced in \cite{Cisneros-Perez2023}. We can generalize the CKM matrix like the product of three rotations
matrices \cite{blanke2007another,blanke2007rare}.

We have considered the case:

\begin{equation}
 V_{Hd}=
 \begin{pmatrix}
  1 & 0 & 0\\
  0 & 1/\sqrt{2} & 1/\sqrt{2}\\
  0 & -1/\sqrt{2} & 1/\sqrt{2}
 \end{pmatrix},
 \end{equation}
\\
 \noindent and through the product $(V_{Hd}^{\dagger})^{-1}V_{CKM}^{\dagger}$, we obtain the matrix $V_{Hu}$ =

\begin{equation}
\label{vhu3}
\begin{aligned}
%a = & \\
& \left(
\begin{matrix}
0.9737 \pm 0.0003 & 0.221 \pm 0.004 & 0.0086 \pm 0.0002 \\
0.227 \pm 0.0008 & 1.0038 \pm 0.0061 & 0.7585 \pm 0.0206 \\
-0.1548 \pm 0.0006 & -0.6486 \pm 0.0045 & 0.9847 \pm 0.0291
\end{matrix}
\right)
\end{aligned}
\end{equation}
where the uncertainties are computed analytically using the reported experimental uncertainties of the CKM-matrix elements.

Finally, to obtain the total contribution of the 2HDM to the chromomagnetic dipole, we sum the scalar and vectorial contributions given by the amplitudes of Eqs.~(\ref{ampE}) and (\ref{ampV}). The computation of $\hat{\mu}_t$ involves several measured boson and quark masses.

\section{Results}
\label{results}
{\color{black}This section presents our results for the $\hat{\mu}_t$ calculations. Fig. \ref{caso3} condense the results for all the studied $\beta$ angles  described in subsection~\ref{Scase3}. In Table~\ref{CMDMS}, we summarize the numerical values of our $\hat{\mu}_t$ calculations along with their uncertainty at 68\% C.L.}

\subsection{\textbf{Case evaluated}}
\label{Scase3}

\begin{figure}[H]
\centering
\includegraphics[width=8.5cm]{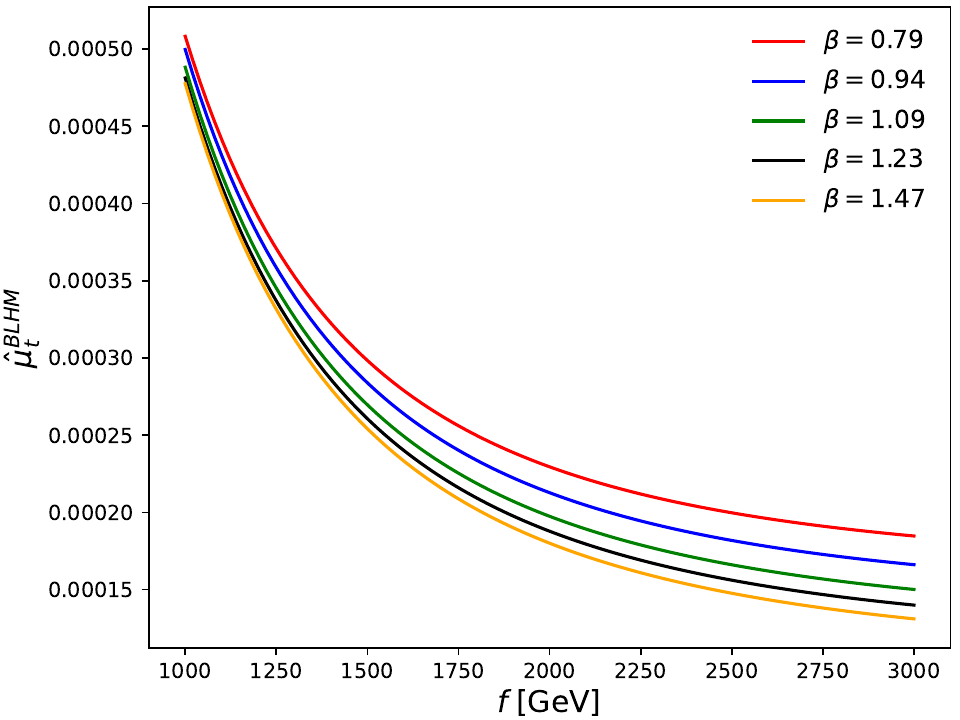}
\caption{\textbf{Case evaluated.} Total contributions to the CMDMs with the CKM matrix of Eq.~(\ref{vhu3}) {\color{black} for the $\beta$ angles considered. The solid lines correspond to the central values of our calculations. }}
\label{caso3}
\end{figure}

In Fig. \ref{caso3}, we show the curves corresponding to the CMDMs with the extended CKM matrix of Eq.~(\ref{vhu3}). The scalar contributions were only positive of order $10^{-6}$, while the vectorial contributions were both positive and negative of order $10^{-4}$.

\begin{table*}
\caption {Numerical values of the CMDM for the case evaluated of the extended CKM matrix. {\color{black} In the first column the $\beta$ angles are indicated. The second column contains the values for the CMDM evaluated at $f=1$ TeV. The third column contains the values for the CMDM evaluated at $f=3$ TeV. The uncertainties are reported at 68\% C.L. }} \label{CMDMS}
\medskip
\resizebox{\columnwidth}{!}{%
\begin{tabular}{c| c | c }\hline \hline

& Case evaluated & Case evaluated \\
\hline

$ \beta $ [rad] & $f=1$ TeV & $f=3$ TeV \\ \hline 
$0.79$ & $(5.08^{+0.13}_{-0.12})\times 10^{-4}$ & $(1.85^{+0.04}_{-0.04})\times 10^{-4}$ \\ 
$0.94$ & $(4.98^{+0.18}_{-0.16})\times 10^{-4}$ & $(1.66^{+0.06}_{-0.06})\times 10^{-4}$ \\ 
1.09 & $(4.87^{+0.21}_{-0.18})\times 10^{-4}$ & $(1.5^{+0.07}_{-0.06})\times 10^{-4}$ \\ 
1.23 & $(4.81^{+0.23}_{-0.20})\times 10^{-4}$ & $(1.39^{+0.08}_{-0.07})\times 10^{-4}$ \\ 
1.47 & $(4.77^{+0.22}_{-0.23})\times 10^{-4}$ & $(1.32^{+0.08}_{-0.07})\times 10^{-4}$ \\ 
\hline \hline

\end{tabular}
}

\end{table*}

Table \ref{CMDMS} displays the numerical values for the CMDM according to $f=1,3$ TeV. 
{\color{black} As previously stated, these values are expected to be nearly identical across all three cases, up to two decimal points. However, differences become apparent when examining values to a greater number of decimal points.}

The CMDM of the SM was computed off-shell in \cite{montano1,montano2} in both its spacelike version $\hat{\mu}_t=-0.0224-0.000925i$ and timelike version $\hat{\mu}_t=-0.0133-0.0267i$. We have also computed the CMDM by taking the off-shell gluon $(q^2\neq0)$ in the two scenarios: the spacelike $(q^2=-m^2_Z)$ and the timelike one $(q^2=m_Z^2)$. In our case, the CMDMs turned out to be identical for both scenarios due to the equality $g_A=g_B$ between the coupling constants of the $SU(2)_A$ and $SU(2)_B$ groups, respectively.

\section{Conclusions}
\label{conclusions}

In this work, we calculate the CMDM for different values of the mixing angle $\beta$, finding values on the order of $10^{-4}$ within the allowed interval for the parameter space.
In this study, we introduced for the first time the extended CKM matrix for the case of chromomagnetic dipole in the BLHM. In the case evaluated, the matrix $V_{Hu}$ given by Eq.~(\ref{vhu3}) also has no important effects on the total CMDMs.
It is important to note that the contributions involving the matrix elements $V_{Hu}$ are very few because we have only quantified the interactions of the top quark and the charged bosons with virtual heavy quarks in the BLHM. That is, only the heavy quarks $B$ and $T^{5/3}$ have intervened on a few occasions.

{
Experimental searches for heavy quarks have found no signals below 2.3 TeV for $T$ and $B$~\cite{ATLAS:2023pja}, and below 1.42 TeV for $T^{5/3},\,T^6,\,T^{2/3}$~\cite{ATLAS:2022tla}. Therefore, studying the CMDM provides an alternative method for searching for these heavy quarks. However, the current experimental precision of the $\hat{\mu}_t$ measurement is as large as 85\%. Theoretically, recent SM calculations for $\hat{\mu}_t$ show differences when comparing the chromodipolar 3-body vertex $gt\bar{t}$ ($\vert\vert\hat{\mu}_{t}^{SM(3b)}\vert\vert=0.0224$)~\cite{montano1, montano2} with the non-abelian 4-body vertex $ggt\bar{t}$ ($\vert\vert\hat{\mu}_{t}^{SM(4b)}\vert\vert=0.0253$)~\cite{tututi2023}. This discrepancy, on the order of $10^{-3}$, might be explained by an extension of the SM, such as the BLHM. The theoretical result presented in this paper is on the order of $10^{-4}$.
}


\begin{thebibliography}{99}
%(1)
    \bibitem{schmaltz2010bestest}
    M.~Schmaltz, D.~Stolarski and J.~Thaler,
    JHEP \textbf{09}, 018 (2010).
%(2)    

%(3)
    \bibitem{Cisneros-Perez2023}
    T. Cisneros-Pérez, M. A. Hernández-Ruíz, A. Gutiérrez-Rodríguez, E. Cruz-Albaro, Eur. Phys. J. C \textbf{83}, 1093 (2023).
%(4)
    \bibitem{kobayashi1973cp}
    M.~Kobayashi and T.~Maskawa,
    %``CP Violation in the Renormalizable Theory of Weak Interaction,''
    Prog. Theor. Phys. \textbf{49}, 652 (1973).
%(5)

%(6)
    \bibitem{Ding:2008nh}
    L.~Ding and C.~X.~Yue,
    %``Top quark chromomagnetic dipole moment in the littlest Higgs model with T-parity,''
    Commun. Theor. Phys. \textbf{50}, 441-444  (2008).
%(7)    

%(8)    

%(9)    

%(10)    

%(11)    

%(12)    

%(13)

%(14)    
    \bibitem{Sirunyan2020}
    A. Sirunyan, {\it et al.}, {\ JHEP} {\bf 06}, 146 (2020).
    
%(15)    

%(16)    

%(17)    
    \bibitem{Hernandez-Juarez:2018uow}
    A.~I.~Hern\'andez-Ju\'arez, A.~Moyotl and G.~Tavares-Velasco,
    %``Chromomagnetic and chromoelectric dipole moments of the top quark in the fourth-generation THDM,''
    Phys. Rev. D \textbf{98}, 035040 (2018).
%(18)    
    \bibitem{Schmaltz:2008vd}
    M.~Schmaltz and J.~Thaler,
    %``Collective Quartics and Dangerous Singlets in Little Higgs,''
    JHEP \textbf{03}, 137 (2009).
%(19)    
    \bibitem{phdthesis}
    Moats, K.   doi: 10.22215/etd/2012-09748, (2012).
%(20)    

%(34)    
    \bibitem{ATLAS:2023pja}
    G.~Aad \textit{et al.} (ATLAS Collaboration),
    %``Search for single production of vector-like T quarks decaying into Ht or Zt in pp collisions at $ \sqrt{s} $ = 13 TeV with the ATLAS detector,''
    JHEP \textbf{08}, 153 (2023).
%(35)    
    \bibitem{ATLAS:2022tla}
    G.~Aad \textit{et al.} (ATLAS Collaboration),
    %``Search for pair-produced vector-like top and bottom partners in events with large missing transverse momentum in pp collisions with the ATLAS detector,''
    Eur. Phys. J. C \textbf{83}, 719 (2023).
%(36)    

%(41)
    \bibitem{blanke2007another}
    M.~Blanke, A.~J.~Buras, A.~Poschenrieder, S.~Recksiegel, C.~Tarantino, S.~Uhlig and A.~Weiler,
    %``Another look at the flavor structure of the littlest Higgs model with T-parity,''
    Phys. Lett. \textbf{B646}, 253 (2007).
%(42)
    \bibitem{blanke2007rare}
    M.~Blanke, A.~J.~Buras, A.~Poschenrieder, S.~Recksiegel, C.~Tarantino, S.~Uhlig and A.~Weiler,
    %``Rare and CP-Violating $K$ and $B$ Decays in the Littlest Higgs Model with $T^-$ Parity,''
    JHEP \textbf{01}, 066 (2007).
%(43)    
    \bibitem{shtabovenko2020feyncalc}
    V.~Shtabovenko, R.~Mertig and F.~Orellana,
    %``FeynCalc 9.3: New features and improvements,''
    Comput. Phys. Commun. \textbf{256}, 107478 (2020).
%(44)
    \bibitem{patel2015package}
    H.~H.~Patel,
    %``Package-X 2.0: A Mathematica package for the analytic calculation of one-loop integrals,''
    Comput. Phys. Commun. \textbf{218}, 66 (2017).
%(45)
    \bibitem{montano1}
    J. I. Aranda, D. Espinosa-Gómez, J. Montaño, B. Quezadas-Vivian, F. Ramírez-Zavaleta and E. S. Tututi, Phys. Rev. D \textbf{98}, 116003 (2018).
%(46)    
    \bibitem{montano2}
    J. I. Aranda, T. Cisneros-Pérez, J. Montaño, B. Quezadas-Vivian, F. Ramírez-Zavaleta and E. S. Tututi, Eur. Phys. J. Plus \textbf{136}, 164 (2021).
%(47)

%(50)
    \bibitem{tututi2023}
    J. Montano-Dom\'inguez, F Ram\'rez-Zavaleta, E. S. Tututi and E. Urquiza-Trejo J. Phys. G: Nucl. Part. Phys. \textbf{50} 115004 (2023).

\end{thebibliography}
\end{document}